\documentclass[aps, prd, superscriptaddress]{revtex4-2}

\usepackage{amsmath,amssymb,bm}
\usepackage[colorlinks]{hyperref}
\usepackage[capitalise]{cleveref}
\usepackage[dvipdfmx]{graphicx}
\usepackage{multirow}

\usepackage{color}
\usepackage{ulem}

\begin{document}

\title{Precision of localization of single gravitational-wave source with pulsar timing array}

\author{Ryo Kato}
\email{rkato@sci.osaka-cu.ac.jp}
\affiliation{Faculty of Advanced Science and Technology, Kumamoto University, Kumamoto 860-8555, Japan}
\affiliation{Osaka Central Advanced Mathematical Institute, Osaka Metropolitan University, Osaka, 5588585, Japan}
\author{Keitaro Takahashi}
\email{keitaro@kumamoto-u.ac.jp}
\affiliation{Faculty of Advanced Science and Technology, Kumamoto University, Kumamoto 860-8555, Japan}
\affiliation{International Research Organization for Advanced Science and Technology, Kumamoto University, Kumamoto 860-8555, Japan
}

\date{\today}

\begin{abstract}
    Pulsar Timing Arrays (PTAs) are expected to be able to detect gravitational waves (GWs) from individual supermassive black hole binaries in the near future.
    In order to identify the host galaxy of a gravitational wave source, the angular resolution of PTAs should be much better than that expected from the conventional methodology of PTAs.
    We study the potential usefulness of precise pulsar-distance measurements in the determination of the sky location of a single GW source.
    Precise distance information from external observations such as astrometry by Very Long Baseline Interferometry is incorporated as priors in the PTA analysis and we evaluate the precision of the sky location of a GW source by simulating PTA data of 12 milli-second pulsars with only the GW signal and the Gaussian white noise in the timing residuals.
    We show that only a few pulsars with a distance precision of 1 pc will improve the precision of the source location by more than 1 order in the presence of white noise of 10 ns.
\end{abstract}

\maketitle

\section{Introduction}\label{sec: Introduction}

A Pulsar Timing Array (PTA) experiment is a method to detect gravitational waves (GWs) in the frequency range of $10^{-9}-10^{-7} {\rm Hz}$ by observing milli-second pulsars over a long period of time \cite{1990ApJ...361..300F}.
There are multiple PTA projects around the world: European Pulsar Timing Array (EPTA) \cite{2023arXiv230616224A}, North American Nanohertz Observatory for Gravitational Waves (NANOGrav) \cite{NANOGrav:2023hde}, Parkes Pulsar Timing Array (PPTA) \cite{2023arXiv230616230Z}, Indian Pulsar Timing Array (InPTA) \cite{Tarafdar:2022toa}, Chinese Pulsar Timing Array (CPTA) \cite{Xu:2023wog} and MeerKAT Pulsar Timing Array (MPTA) \cite{Miles:2022lkg}.
International Pulsar Timing Array (IPTA) is a consortium that coordinates and enhances the activities of these PTAs and combines data from different PTAs \cite{Antoniadis:2022pcn}.
The primary GW source for the PTAs is supermassive black hole binaries (SMBHBs), while GWs from cosmic strings and cosmic inflation are also intriguing targets.

Recently, non-negligible Bayesian evidence for stochastic GW background has been reported by EPTA+InPTA, NANOGrav, PPTA and CPTA \cite{Reardon:2023gzh,2023arXiv230616214A,NANOGrav:2023gor,Xu:2023wog}.
Although the current data and analysis are not yet statistically significant enough to claim detection, in the near future, by further accumulating data and combining data from different PTAs, it will be determined whether the obtained signal originates from the stochastic GW background.
Then one of the next goals of PTAs is to detect continuous GWs from individual SMBHBs \cite{Rosado:2015epa}.
If we can detect and precisely locate single GW sources, we will be able to identify the host galaxies of the SMBHBs and, in the future, create an SMBHB catalog.
This is nanohertz GW astronomy, and follow-up observations with electromagnetic waves will enable us to study the evolution and growth of SMBHs from multiple aspects.

As an example, let us consider an SMBHB with a chirp mass of $10^{9} ~M_{\odot}$ and GW frequency of $10^{-8}~{\rm Hz}$.
If it is located within about $1~{\rm Gpc}$ ($z=0.2$) from the observer, the GW amplitude is about $3 \times 10^{-8}~{\rm s}$ and detectable with the current PTAs in the near future.
On the other hand, according to the Galaxy And Mass Assembly survey, the surface number density of galaxies below the redshift $z=0.2$ is about $10^3~{\rm deg^{-2}}$ \cite{Baldry_2010,Driver:2022vyh}.
Therefore, the host galaxy of a GW source would not be identified without the resolution of the GW source of about $10^{-3}~{\rm deg^2} \approx 4~{\rm amin^2}$.
So far, many researchers have studied the precision of the sky location determination for a continuous GW source \cite{Sesana:2010ac,2010arXiv1008.1782C,Lee:2011et,Babak_2012,Ellis:2013hna,Taylor:2014iua,Wang:2014ava,Taylor:2015kpa,wang2016coherent,Zhu:2016zlx,Wang:2016tfy,Wang:2020hfh,Goldstein_2018,Chen:2022ooe,Songsheng:2021tri,becsyFastBayesianAnalysis2022}.
According to simulations of the IPTA dataset, the sky location can be determined with a precision on the order of 100 $\rm deg^2$ \cite{Goldstein_2018}.
This precision will not be sufficient to identify the host galaxy among a large number of candidate galaxies.

In this paper, we study the potential usefulness of precise pulsar-distance measurements in the determination of the sky location of a single GW source.
In the GW signal model, the pulsar distance appears mainly in the phase of the "pulsar term".
Precise pulsar distances from external observations such as VLBI astrometry can be incorporated as priors in the PTA analysis and are expected to improve the determination of parameters including the sky location of the GW source.
Given that a typical period of GWs targeted by PTAs is O(1-10) years, the phase of the pulsar term can be determined only if the uncertainty in the pulsar distance is less than 1 pc.

There are several different ways to measure pulsar distances.
Among these, pulsar distances obtained with Very Long Baseline Interferometry (VLBI) astrometry are independent of the timing model used by the PTAs and can be used as prior information for the PTA analysis \cite{Ding:2022luk}.
For example, the Very Long Baseline Arrays (VLBA) have been used to measure the parallax-based pulsar distance of many pulsars \cite{dellerMicroarcsecondVLBIPulsar2019}.
Recently, milli-second pulsars have also been observed with the VLBA for distance measurements \cite{Ding:2022luk}.
According to the VLBA results, for example, the parallax of PSR J0030+0451 was measured to be 3.02(7) mas and the distance was determined to be $331(8)$ pc without a timing model.
In the SKA (Square Kilometre Array) era, combining the SKA with other radio telescopes to perform VLBI observations, the maximum astrometric precision is expected to reach 15 $\mu$as at 1.4 GHz \cite{refId0}.
With this precision, the uncertainty in the distance of a pulsar located at 300 pc will be on the order of 1 pc.

In this paper, we evaluate the potential effects of precise pulsar distance on the determination of PTA parameters, especially the location of a GW source, considering the current and future precision of pulsar distance measurements.
The paper is organized as follows.
In \cref{sec: Method}, we describe the signal model of a GW emitted from a circular SMBHB and the parameter estimation method.
In \cref{sec: Results}, we investigate the results of the analysis for simulated data and compare the results with several models.
In \cref{sec: Conclusion}, we provide our conclusions and discuss future work.
Throughout this paper, we use units where $G = c = 1$.

\section{Method}\label{sec: Method}

\subsection{Signal Model}\label{subsec: Signal Model}

In General Relativity, a GW is represented as a superposition of the plus ($+$) and cross ($\times$) polarizations:
\begin{align}
    h_{\mu\nu}(t,\hat{\Omega})=\sum_{A=+,\times}h_{A}(t,\hat{\Omega})e_{\mu\nu}^{A}(\hat{\Omega}),
\end{align}
where $\hat{\Omega}$ is a unit vector from the GW source to the Solar System Barycenter (SSB), $h_{A}$ are the polarization amplitudes and $e_{\mu\nu}^{A}$ are the polarization tensors.
The polarization tensors can be expressed as
\begin{align}
    e_{\mu \nu}^{+}(\hat{\Omega})=\hat{m}_{\mu}\hat{m}_{\nu}-\hat{n}_{\mu}\hat{n}_{\nu}, \\
    e_{\mu \nu}^{\times}(\vec{\Omega})=\hat{m}_{\mu}\hat{n}_{\nu}+\hat{n}_{\mu}\hat{m}_{\nu},
\end{align}
where $\hat{m}$ and $\hat{n}$ are unit vectors orthogonal to each other and to $\hat{\Omega}$.

The GW induces the timing residuals \cite{anholmOptimalStrategiesGravitational2009,finnResponseInterferometricGravitational2009,bookAstrometricEffectsStochastic2011}:
\begin{align}
    s(t,\hat{\Omega})=\sum_{A=+,\times}\Delta s_{A}(t)F^{A}(\hat{\Omega}),\label{eq:signal}
\end{align}
where
\begin{align}
    \Delta s_{A}(t)=s_{A}(t_{p})-s_{A}(t).
    \label{eq:residual}
\end{align}
Here, $t$ and $t_{p}$ are times when the GW passed through the SSB and the pulsar, respectively, and $F^{A}$ are the antenna pattern functions.
The first and second terms in Eq.~(\ref{eq:residual}) are called pulsar and Earth terms, respectively.
Let $\hat{p}$ be a unit vector from the SSB to the pulsar, then the time $t_{p}$ is written as
\begin{align}
    t_{p}=t-L_{p}(1+{\hat{\Omega}}\cdot{\hat{p}}),
    \label{eq:tp}
\end{align}
where $L_{p}$ is the distance between the SSB and the pulsar.
Furthermore, the antenna pattern functions are defined as,
\begin{align}
    F^{A}({\hat{\Omega}})\equiv{\frac{1}{2}}{\frac{{\hat{p}}^{\mu}{\hat{p}}^{\nu}}{1+{\hat{\Omega}}\cdot{\hat{p}}}}e_{\mu\nu}^{A}({\hat{\Omega}}).
\end{align}
We define the unit vectors as
\begin{align}
    \hat{\Omega} & =-(\sin\theta\cos\phi)\hat{x}-(\sin\theta\sin\phi)\hat{y}-(\cos\theta)\hat{z},                    \\
    \hat{m}      & =(\sin\phi)\hat{x}-(\cos\phi)\hat{y},                                                             \\
    \hat{n}      & =-(\cos\theta\cos\phi)\hat{x}-(\cos\theta\sin\phi)\hat{y}+(\sin\theta)\hat{z},                    \\
    \hat{p}      & =(\sin\theta_{p}\cos\phi_{p})\hat{x}+(\sin\theta_{p}\sin\phi_{p})\hat{y}+(\cos\theta_{p})\hat{z},
\end{align}
where ${\hat{x}},{\hat{y}}$ and ${\hat{z}}$ are the Cartesian basis vectors and $\hat{\Omega}=\hat{m}\times\hat{n}$.
In the case of a GW emitted from a circular SMBHB, $s_{A}$ can be written as \cite{Ellis:2014xgh}:
\begin{align}
    s_{+}(t)      & =\frac{{\mathcal M}^{5/3}}{d_{L}\omega_{s}(t)^{1/3}}\left[\sin2\Phi_{s}(t)\left(1+\cos^{2}\iota\right)\cos2\psi+2\cos2\Phi_{s}(t)\cos\iota\sin2\psi\right],          \\
    s_{\times}(t) & =\frac{\mathcal{M}^{\mathrm{5/3}}}{d_{L}\omega_{s}(t)^{1/3}}\left[-\sin2\Phi_{s}(t)\left(1+\cos^{2}\iota\right)\sin2\psi+2\cos2\Phi_{s}(t)\cos\iota\cos2\psi\right],
\end{align}
where $\mathcal{M}\equiv(m_{1}m_{2})^{3/5}/(m_{1}+m_{2})^{1/5}$ is the chirp mass of the SMBHB with the individual black hole masses $m_{1}$ and $m_{2}$, $d_{L}$ is the luminosity distance of the SMBHB, $\iota$ is the inclination angle of the SMBHB and $\psi$ is the GW polarization angle.
The orbital angular frequency and the phase of the SMBHB are
\begin{align}
    \omega_{s}(t) &
    =\omega_{s0}\left(1-\frac{256}{5}{\mathcal M}^{5/3}\omega_{s0}^{8/3}t\right)^{-3/8}
    =\omega_{s0}\left(1-\frac{t}{t_{\rm coal}}\right)^{-3/8}, \label{eq:omega} \\
    \Phi_{s}(t)   &
    =\Phi_{s0}+{\frac{1}{32{\mathcal M}^{5/3}}}\left(\omega_{s0}^{-5/3}-\omega_{s}(t)^{-5/3}\right)
    =\Phi_{s0}+\frac{8}{5} \omega_{s0} t_{\rm coal} \left[1-\left(1-\frac{t}{t_{\rm coal}}\right)^{5/8}\right],
\end{align}
where $\omega_{s0}=2\pi f_{s0}$ and $\Phi_{s0}$ are the initial values of the orbital angular frequency and phase at $t=0$, respectively.
Here, the coalescing time of the SMBHB, at which the orbital angular frequency $\omega_{s}$ diverges, is defined as
\begin{align}
    t_{\rm coal}=\frac{5}{256}\mathcal{M}^{-5/3}\omega_{s0}^{-8/3}.
\end{align}
The initial GW frequency and phase are related to these quantities as $f_{0}=2f_{s0}$ and $\Phi_{0}=2\Phi_{s0}$, since the GW frequency and phase are twice the orbital frequency and phase of SMBHB, respectively.

It should be noted that, when $t \ll t_{\rm coal}$, as is the normal situation during the observation period of the PTA, $\omega_{s}(t)$ is constant and the phase is reduced to,
\begin{equation}
    \Phi_{s}(t) = \Phi_{s0} + \omega_{s0} t.
    \label{eq:phase_earth}
\end{equation}
On the other hand, orbital angular frequency in the pulsar term is, for $t \ll t_{\rm coal}$,
\begin{equation}
    \omega_{s}(t_p) = \omega_{s0}\left( 1 + \frac{L_p(1+{\hat{\Omega}}\cdot{\hat{p}})}{t_{\rm coal}} \right)^{-3/8}.
    \label{eq:omega_p}
\end{equation}
If the value of $L_p(1+{\hat{\Omega}}\cdot{\hat{p}})$ is not negligible compared to $t_{\rm coal}$, the angular frequency of the pulsar term is smaller than that of Earth term.
Further, the phase of the pulsar term is given by,
\begin{equation}
    \Phi_{s}(t_p)
    = \Phi_{s0}
    + \frac{8}{5} \omega_{s0} t_{\rm coal} \left[1-\left(1+\frac{L_p(1+{\hat{\Omega}}\cdot{\hat{p}})}{t_{\rm coal}}\right)^{5/8}\right]
    + \omega_{s}(t_p) t.
    \label{eq:phase_p}
\end{equation}
Therefore, compared to Eq.~(\ref{eq:phase_earth}), there is an offset in the initial phase which depends on the value of $L_p(1+{\hat{\Omega}}\cdot{\hat{p}})$, the chirp mass and angular frequency.

\subsection{Simulated data}\label{subsec: Simulated data}

\begin{figure}[tb]
    \centering
    \includegraphics[width=18cm]{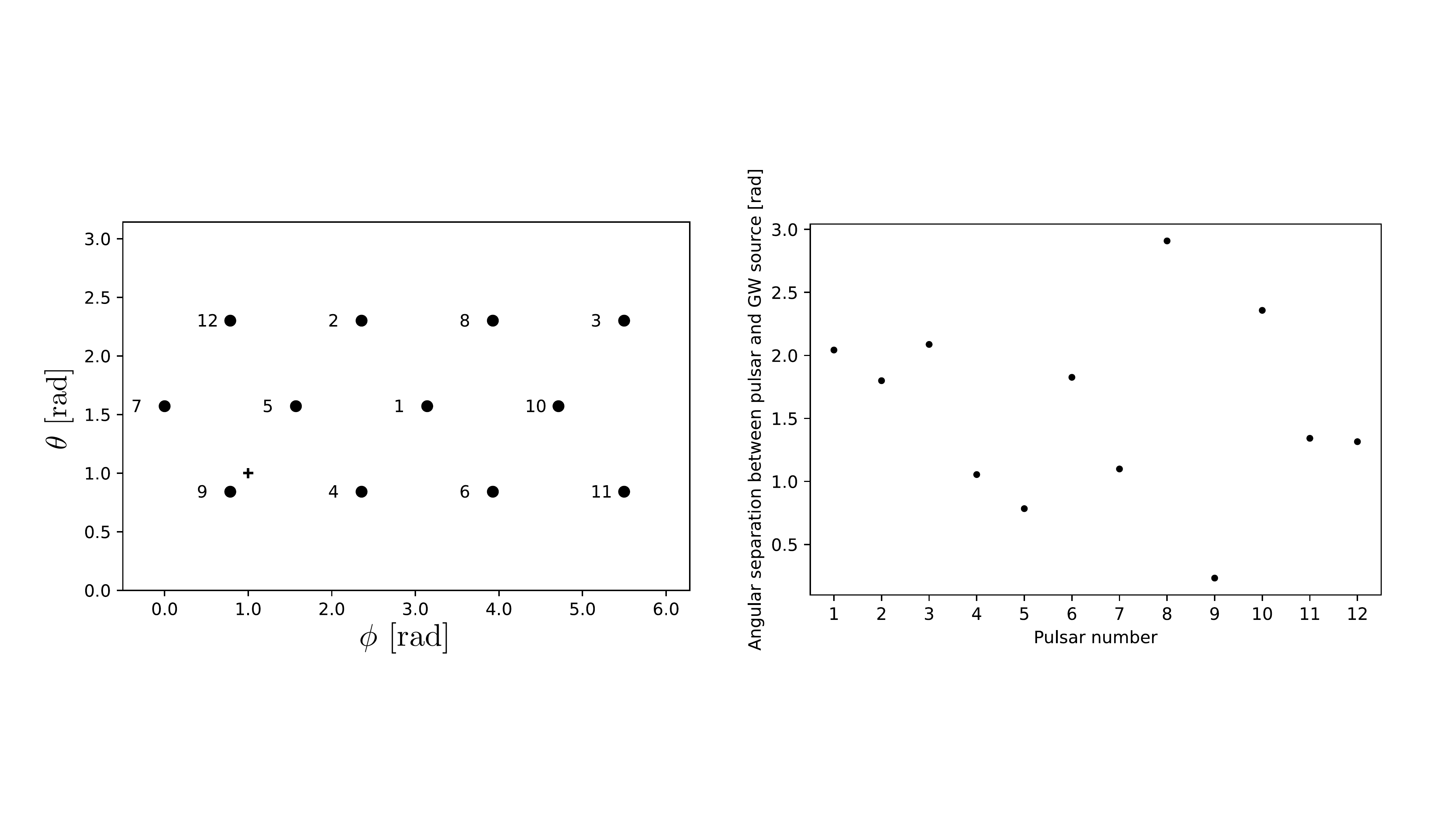}
    \caption{
        {\it Left}: Sky locations of the pulsars and the GW source.
        The dot and the plus markers denote the Sky locations of the pulsars and the GW source.
        The numbers to the left of the dot markers indicate the numbers randomly assigned to the pulsars.
            {\it Right}: Angular separation between each pulsar and the GW source.
        The pulsar number is given in the left panel.
    }
    \label{fig:posdist}
\end{figure}\par

In our simulation, the timing residual is evenly sampled once every three weeks with an observation period of 12.5 years.
We consider 12 pulsars distributed uniformly in the sky.
\cref{fig:posdist} shows the sky location of the 12 pulsars and the GW source in the left panel and shows angular separation between each pulsar and the GW source in the right panel.
The distance from the SSB is set to 1 kpc for all pulsars.
The pulsar distance is one of the key ingredients in our study and will be mentioned later in \cref{subsec: Pulsar Distance Prior}.

The simulated timing residual consists of the GW signal of a circular SMBHB, whose sky location is given in \cref{fig:posdist}, and the Gaussian white noise.
The GW signal parameters are set to $\theta = 1 $, $\phi = 1$, $\mathcal M=10^9\, M_{\odot}$, $d_{L}=10^2\, {\rm Mpc}$, $f_{0}=10^{-8}\, {\rm Hz}$, $\Phi_{0}=1$, $\iota=1$ and $\psi=1$.
In this case, the coalescing time scale $t_{\rm coal}$ is about $7000\, {\rm yr}$.
This time scale is sufficiently longer than the observation period so that the evolution of the binary can be neglected during the observation period.
On the other hand, the coalescing time is longer than the light travel time between the SSB and the pulsars by a factor of 2.
Thus, according to Eq.(\ref{eq:omega}), the minimum GW frequency of the pulsar terms is 0.8 times that of the Earth term.
With the above set of parameters, the GW signal has an amplitude of approximately 10 ns.
Then, the standard deviation of the Gaussian white noise is set to either of $1 \rm{ns}$, $10 \rm{ns}$ and $100 \rm{ns}$.
The same realization of the noise was used in all of the analyses, although the magnitude was different.

We define the simulated timing residual data $\delta t$ which consists of the GW signal $s$ given by \cref{eq:signal} and the Gaussian white noise $n$ as,
\begin{align}
    \delta t = s + n.
\end{align}
We do not take into account the timing model error and possible time-correlated noise such as red noise.

\subsection{Parameter Estimation}\label{subsec: Parameter Estimation}

In the Bayesian framework, given a parameterized model $M$ that describes the data $\delta t$ with parameters $\theta$, the posterior probability density function (PDF) is \cite{gregory_2005}
\begin{align}
    p(\theta|\delta t,M)=\frac{p(\delta t|\theta,M)p(\theta|M)}{p(\delta t|M)},
\end{align}
where $p(\delta t|\theta,M)$ is the likelihood function, $p(\theta|M)$ is the prior PDF and $p(\delta t|M)$ is an uninteresting normalization constant.
The purpose of the Bayesian parameter estimation is to obtain the posterior PDF.

Given a model $M$, which is $\delta t = s + n$, the likelihood function is given by
\begin{align}
    p(\delta t|\theta,M)=\frac{1}{\sqrt{2\pi C}}\exp\left(-\frac{1}{2}(\delta t-s)^{T}C^{-1}(\delta t-s)\right),
\end{align}
where $\theta$ denotes the parameters for the GW signal model and the noise, and $C \equiv \langle n n^T \rangle$ is the noise covariance matrix.
We assume that the covariance matrix $C$ is proportional to the identity matrix as
\begin{align}
    C=\sigma_{n} I,
\end{align}
where $\sigma_{n}$ is a constant and $I$ is the identity matrix because we do not include time-correlated noise which has a non-diagonal covariance matrix.
Assuming that the pulsar sky location $(\theta_{p}, \phi_{p})$ is determined with high precision for all pulsars, we look for best-fit parameters and obtain the posterior PDF by multiplying the prior PDFs $p(\theta|M)$ summarized in \cref{Prior distributions}.
Specifically, we analyze simulated data using {\tt DEMetropolisZ}, which is the Markov chain Monte Carlo (MCMC) method in {\tt PyMC} package \cite{thomas_wiecki_2022_6396757}.
This method is based on Adaptive Differential Evolution Metropolis \cite{terbraakDifferentialEvolutionMarkov2008}.
For all analyses, we run 4 parallel chains with $2\times10^6$ iterations and discard the first $10^6$ iterations as burn-in.

\subsection{Pulsar Distance Prior}\label{subsec: Pulsar Distance Prior}

\begin{table}[tb] 
    \caption{Prior distributions.}
    \label{Prior distributions}
    \centering
    \normalsize
    \begin{center}
        \begin{tabular}{lll}
            \hline
            Parameter                  & Description         & Prior                           \\
            \hline
            GW signal                                                                          \\
            $\theta$ [rad]             & Polar angle         & Uniform$[0,\pi]$                \\
            $\phi$ [rad]               & Azimuthal angle     & Uniform$[0,2\pi]$               \\
            $\iota$ [rad]              & Inclination angle   & Uniform$[0,\pi]$                \\
            $\mathcal M$ [$M_{\odot}$] & Chirp mass          & Log-uniform$[10^{6},10^{10}]$   \\
            $f_{0}$ [Hz]               & Initial Frequency   & Log-uniform$[10^{-9},10^{-7}]$  \\
            $d_{L}$ [Mpc]              & Luminosity distance & Log-uniform$[10^{0},10^{3}]$    \\
            $\Phi_{0}$ [rad]           & Initial Phase       & Uniform$[0,2\pi]$               \\
            $\psi$ [rad]               & Polarization angle  & Uniform$[0,\pi]$                \\
            $L_{p}$ [pc]               & Pulsar distance     & Normal$(1000,\sigma_{p})$       \\
            \hline
            White noise                                                                        \\
            $\sigma_{n}$ [s]           & Standard deviation  & Log-uniform$[10^{-11},10^{-5}]$ \\
            \hline
        \end{tabular}
    \end{center}
\end{table}

In this paper, as mentioned before, we assume that the distance to all pulsars is 1 kpc and that we have external information on it from independent observations such as VLBI astrometry.
The external information can be incorporated as the prior for pulsar distances, Normal$(1000,\sigma_{p})$, where $\sigma_{p}$ represents 1-$\sigma$ error of the independent observations in the unit of pc, while the mean of the prior is set to the true value of 1 kpc.

Since the distance of pulsars is usually not determined with high precision, the phase of the pulsar term is not determined and should be marginalized with a uniform prior.
However, if the pulsar distance and then the phase of the pulsar term are determined precisely, the determination of other parameters is expected to improve.
More specifically, the phase of the pulsar term is well determined if the uncertainty in the pulsar distance is much smaller than the GW wavelength.
Usually, PTAs are targeted at GWs with a wavelength of about 1pc-10pc, so in order to determine the phase of the pulsar term precisely, the distance of the pulsar must be determined with precision better than about 1pc.

At present, PSR J0030+0451 is the PTA pulsar with the most precisely determined distance, and its precision is better than 10 pc.
In the future, the SKA-VLBI, a combination of the SKA and other facilities, will have much better precision and we can expect 1 pc or smaller distance errors for bright and close pulsars.
Given these facts, we consider several different values of $\sigma_{p}$, reflecting the near and future prospects for precise determination of pulsar distance.
Specifically, we consider values of 100 pc, 10 pc, 1 pc, 0.1 pc and 0.01 pc for $\sigma_{p}$.
The current typical precision of pulsar distance is about 100 pc and we take this as the fiducial value.
In this case, the distance prior will not have any significant effects on the parameter determination.
On the other hand, the precision of 0.01 pc is unrealistic even with the SKA-VLBI.
We consider this extreme case to demonstrate the ultimate potential of the determination of the pulsar distance.

In fact, in determining the pulsar distance by annual parallax measurement, the distance of a closer pulsar tends to be determined more precisely.
However, the determination of the phase of the pulsar term involves the absolute error, rather than the relative error.
In this paper, we put all pulsars at the distance of 1 kpc for convenience, but because the crucial factor is the absolute precision of the pulsar distances, this assumption will not affect the results significantly.

\section{Result}\label{sec: Results}

\begin{figure}[tb]
    \centering
    \includegraphics[width=18cm]{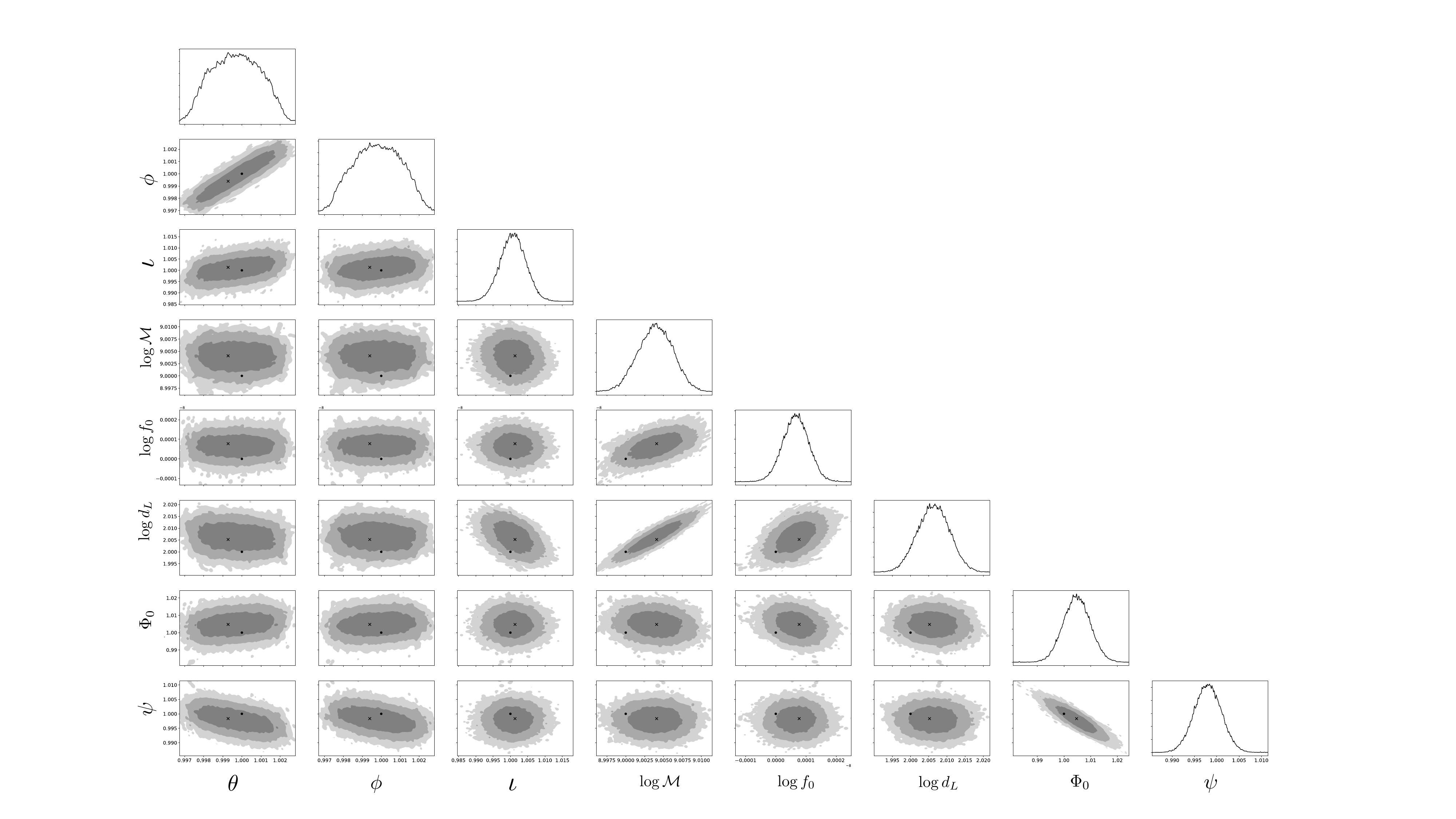}
    \caption{Posterior of the parameters except for the pulsar distances using $\sigma_{p}=100$ pc.
        Contour lines represent 1-$\sigma$, 2-$\sigma$ and 3-$\sigma$ regions.
        The dot and the cross markers represent the true values and the maximum a posteriori (MAP) values, respectively.
    }
    \label{fig:2dpost_1-1}
\end{figure}\par

In general, the structure of the likelihood function in parameter space is complicated and it is expected that there are many local maxima.
In such a case, it is not easy for an MCMC chain to reach the global maximum starting from a random initial point in parameter space.
In this paper, we set the initial point to the global maximum in order to see the effect of the precise distance of pulsars on the precision of the position determination of a gravitational-wave source.
We measure the size of the expected confidence region by examining the structure of the likelihood function around the global maximum.

First, we consider a case with a conservative precision in pulsar distances of 100 pc, while we take unrealistically small white noise of 1 ns.
\cref{fig:2dpost_1-1} shows the corner plot of the posteriors of the GW source parameters.
Since the wavelength of the GW is assumed to be about 3 pc, the phase of the pulsar term would not be restricted at all with this precision of pulsar distance.
The angular resolution of the GW source, the uncertainty in $\theta$ and $\phi$, is about $2 \times 10^{-3}~{\rm rad} \approx 7~{\rm amin}$, and the uncertainty area is about $40~{\rm amin}^2$ for 1-$\sigma$.
This uncertainty is much smaller than those found in the literature due to the extremely small white noise.
As we see, the position parameters, $\theta$ and $\phi$, are not correlated with the other parameters, while they are correlated with each other.
Several parameter pairs, such as $f_{0}$-$\mathcal{M}$, $d_{L}$-$\mathcal{M}$ and $\Phi_{0}$-$\psi$, are also seen to be correlated.

\begin{figure}[tb]
    \centering
    \includegraphics[width=18cm]{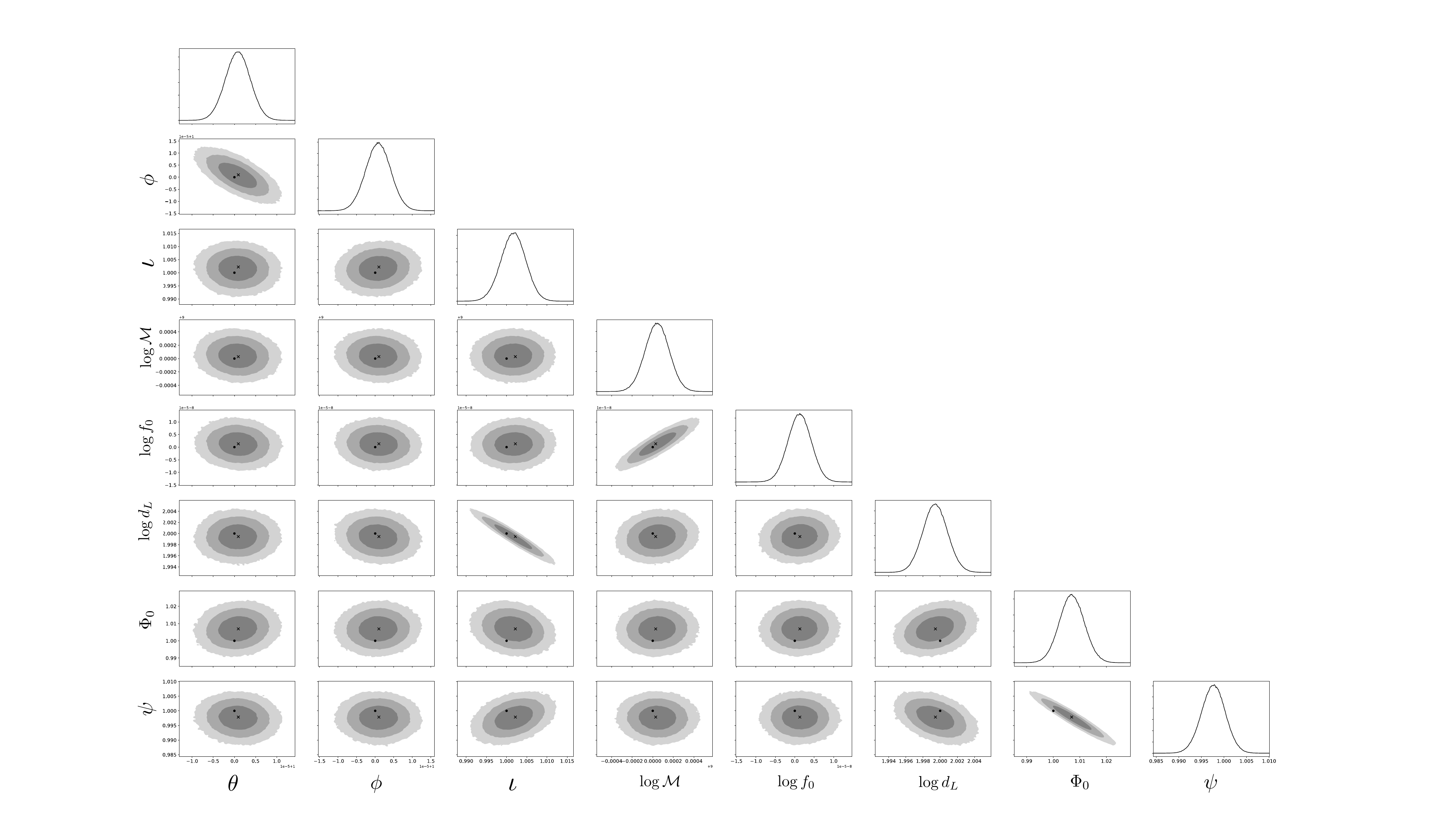}
    \caption{Same as \cref{fig:2dpost_1-1}, but $\sigma_{p}=0.01$ pc is assumed.
    }
    \label{fig:2dpost_1-5}
\end{figure}

Next, we consider a case with 0.01 pc for the precision of pulsar distances, keeping the white noise level of 1 ns.
In contrast to the previous case, with this extreme precision, the phase of the pulsar term would be determined very precisely.
\cref{fig:2dpost_1-5} shows the posteriors of the GW source parameters.
We see that, compared with \cref{fig:2dpost_1-1}, the precision of some of the parameters such as $\theta$, $\phi$, $\mathcal{M}$ and $f_{0}$ is improved by more than 1 order of magnitude.
These parameters are directly related to the phase of the pulsar term, as we saw in Eqs. (\ref{eq:omega_p}) and (\ref{eq:phase_p}).
In particular, the angular resolution of the GW source improves by 2 orders and is about $10^{-5}~{\rm rad} \approx 2~{\rm asec}$, and the uncertainty area is about $2~{\rm asec}^2$ for 1-$\sigma$.
Since the phase of the pulsar term is sufficiently constrained with this precision of pulsar distance, even better precision will not lead to a better angular resolution.
On the other hand, significant improvement cannot be seen in other parameters such as $\iota$, $d_{L}$, $\Phi_{0}$, $\psi$ and $\sigma_n$, which are not directly related to the phase of the pulsar term.
The position parameters, $\theta$ and $\phi$, are correlated again but the direction of correlation is different.
Parameter pairs, such as $f_{0}$-$\mathcal{M}$, $d_{L}$-$\iota$ and $\Phi_{0}$-$\psi$, are also seen to be correlated.

\par

Let us focus more on the precision in the sky location of the GW source.
As we saw above, the precision ideally reaches $2~{\rm asec}^2$, while it is $40~{\rm amin}^2$ for more practical distance errors of 100 pc.
Our simulation assumes that all pulsars are at a distance of 1 kpc and have the same distance error.
In reality, however, pulsars are located at various distances and the distance errors are also different.
In general, pulsars at relatively short distances tend to have small errors for the same precision of parallax measurements.
Since the precision of the phase of the pulsar term is related not to the relative precision of the pulsar distance but to the absolute precision, if the distances of several nearby pulsars are precisely determined, the phase of their pulsar term is also precisely determined.
To simulate such a more practical situation, we consider a case in which the distance of only some of the 12 pulsars is precisely measured with a precision of 0.01 pc, while others have a precision of 100 pc.

\begin{figure}[tb]
    \centering
    \includegraphics[width=18cm]{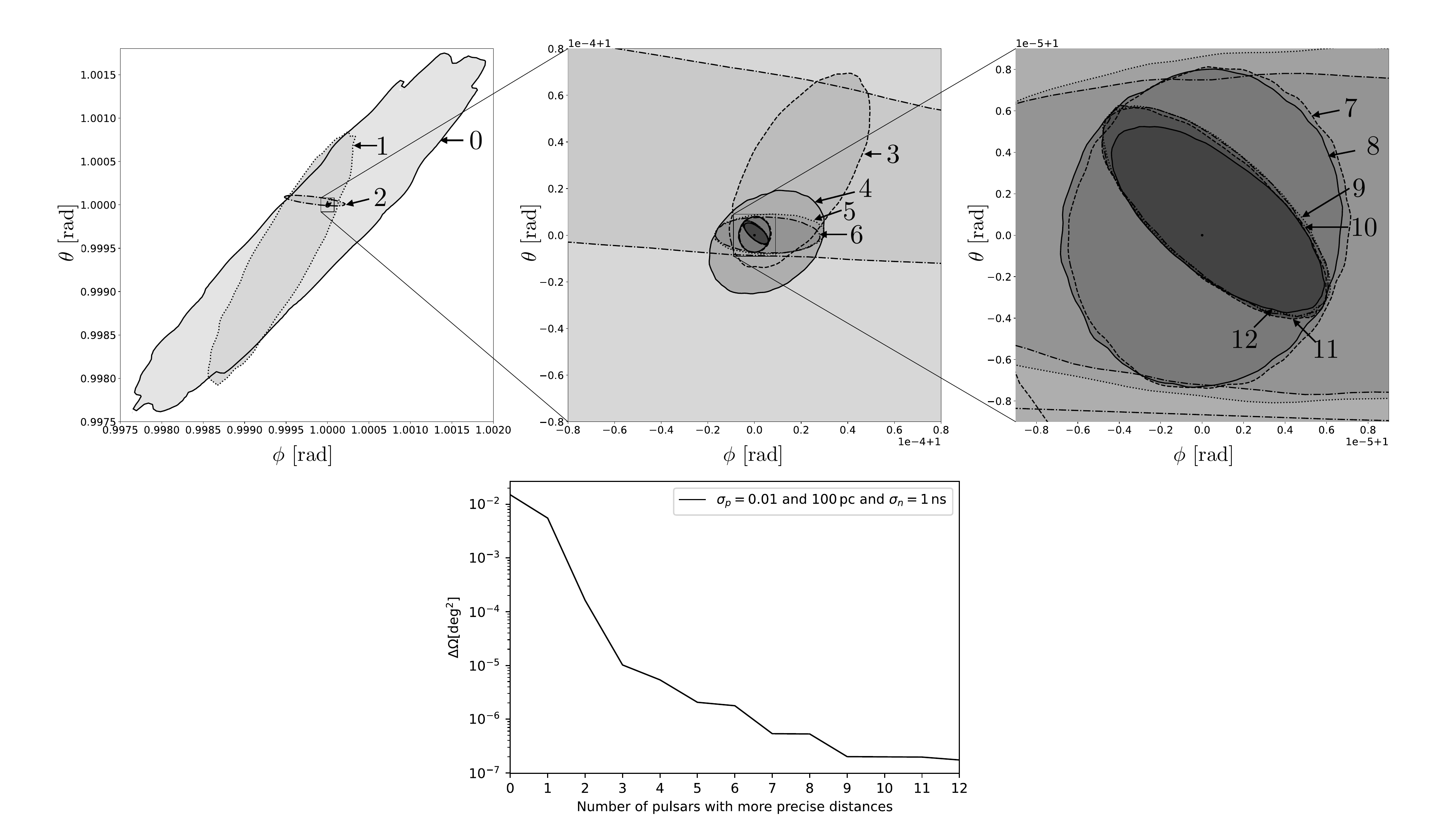}
    \caption{
        {\it Upper}: Posterior of the sky location of SMBHB with 1 ns white noise.
        The contours represent 1-$\sigma$ regions.
        The number of the contours corresponds to the number of pulsars with precisely determined distances.
        We used $\sigma_{p}=100$ pc for pulsars with imprecise distance and $\sigma_{p}=1$ pc for pulsars with precise distance.
        The dot marker represents the true value of the sky location of SMBHB.
            {\it Lower}: Uncertainty of the sky location as a function of the number of pulsars with precisely determined distances.
        The vertical axis is the area of the contour line in the upper panel.}
    \label{fig:tp}
\end{figure}\par

In the upper panel of \cref{fig:tp}, the 1-$\sigma$ uncertainty of the sky location ($\theta$ and $\phi$) with 1 ns white noise is shown for different numbers of pulsars with a precise distance.
The largest and smallest contours are the same as those in \cref{fig:2dpost_1-1} and \cref{fig:2dpost_1-5}, respectively.
Here, in all analyses, we will give a precise distance to pulsars in the ascending order of the assigned number in \cref{fig:posdist}.
It is seen that the area of uncertainty is reduced drastically as the number increases.
Further, the direction of the correlation between $\theta$ and $\phi$ changes as the number increases and this behavior is dependent on the order of the pulsars given the precise distance.
We will briefly discuss the direction of the correlation in \cref{Appendix direction} and focus on the area of the uncertainty region hereafter.

In the lower panel of \cref{fig:tp}, we show the area of the 1-$\sigma$ uncertainty regions shown in the left panel as a function of the number of pulsars with a precise distance.
The precision of the source location estimation improves as the number increases.
It is noticeable that only 3 pulsars with a precise distance drastically improve the source localization.
The precision is saturated to $2 \times 10^{-7}~{\rm deg}^2$ with 9 pulsars.
Thus, it is implied that only a few pulsars with a precise distance can have a significant impact.
Moreover, as shown in the right panel of \cref{fig:posdist}, the first 3 pulsars are located relatively far from the GW source.
In fact, when the closest pulsar (No.9) is given a precise distance, the improvement of the angular resolution is modest compared to the first 3 pulsars.
Thus, it is implied that the proximity of pulsars with a precise distance to the GW source may not be essential.

Let us see the impact of precise pulsar distance measurement more systematically changing the precision of the distances and the white-noise levels.
This allows us to assess how much angular resolution for GW sources we can have in the near and far future.

\begin{figure}[tb]
    \centering
    \includegraphics[width=18cm]{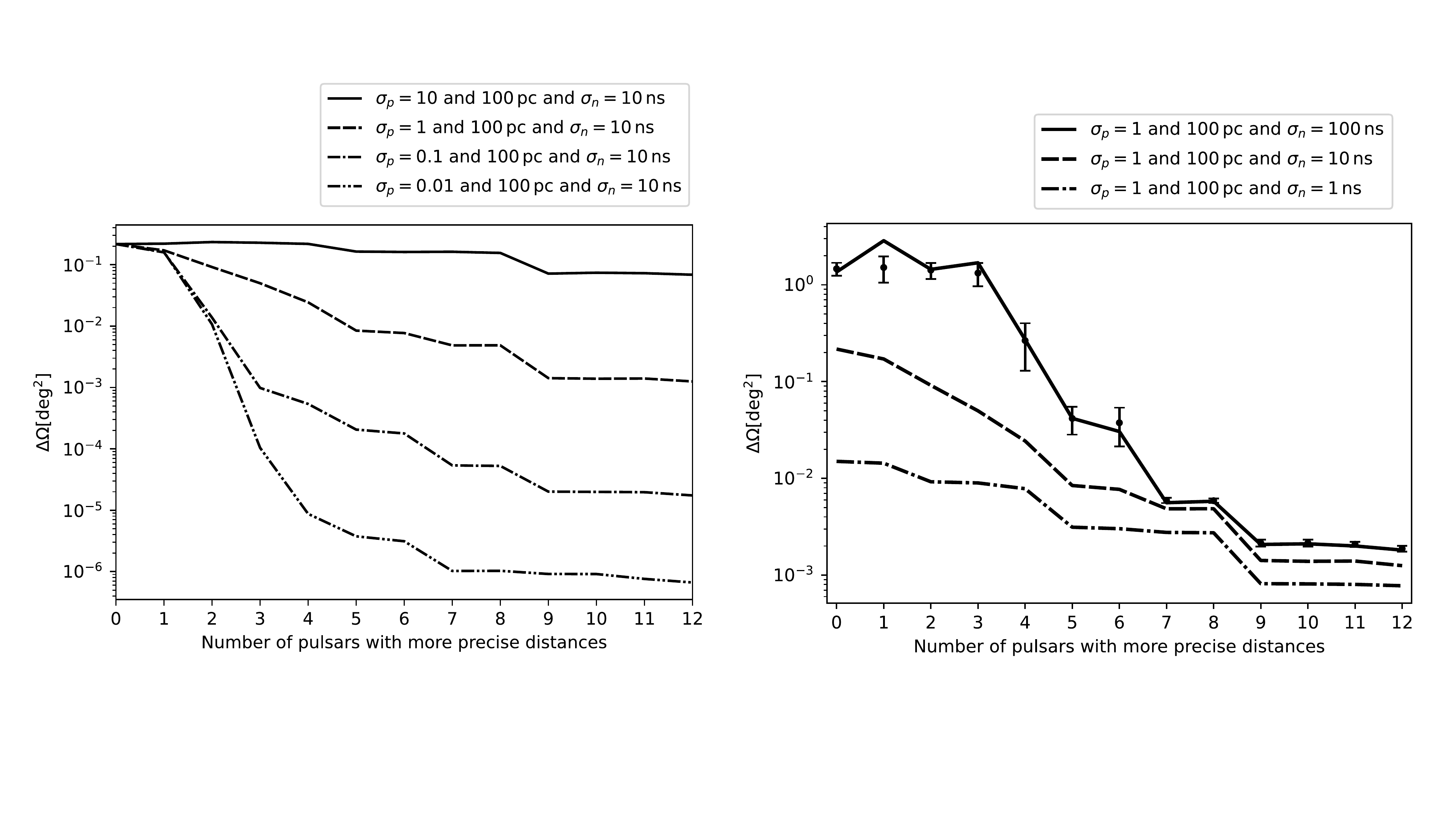}
    \vspace{-2cm}
    \caption{{\it Left}: Same as the right panel of \cref{fig:tp} but with different precision of pulsar distance, $\sigma_{p}=10$ pc (solid), $\sigma_{p}=1$ pc (dashed), $\sigma_{p}=0.1$ pc (dashed-dotted) and $\sigma_{p}=0.01$ pc (dashed double-dotted), while fixing the white-noise level to 10 ns.
            {\it Right}: Same as the right panel of \cref{fig:tp} but with different white-noise levels, $\sigma_{n}=1$ ns (dashed-dotted), $\sigma_{n}=10$ ns (dashed) and $\sigma_{n}=100$ ns (solid), while fixing the distance precision to 1 pc.
        The dots and error bars represent the means and standard deviations of the results using different white-noise realizations.
    }
    \label{fig:deg2}
\end{figure}

First, we compare 4 cases with different values of distance uncertainty (0.01 pc, 0.1 pc, 1 pc and 10 pc), while fixing the white-noise level to 10 ns, which will be practical in the SKA era.
We again give the above precise distance to some of the 12 pulsars, while other pulsars have a distance uncertainty of 100 pc.
The left panel of \cref{fig:deg2} shows the 1-$\sigma$ area of the source localization.
Since the noise level is higher than in the previous cases, the angular resolution is generally about 1 order of magnitude worse for 0.01 pc.
We see that the influence of the distance precision is significant.
When the pulsar distance error increases by one order, the best angular resolution (the rightmost value) deteriorates by 1.5 orders of magnitude.
Specifically, if the distance error is 10pc, the improvement is at most only about a factor of 2 compared to the case of 100 pc.
This is because the phase of the pulsar term is not well determined with an error of 10 pc.
If all pulsars have a distance error of 0.1-1 pc, which will be reachable in the SKA era, the area of the confidence region is about $3 \times 10^{-5}$ - $3 \times 10^{-3}~{\rm deg}^2 = 0.1$ - $10~{\rm amin}^2$.
Even if only 5 out of 12 pulsars have such precise distance, the area is as small as $3 \times 10^{-4}$ - $10^{-2}~{\rm deg}^2 = 1$ - $30~{\rm amin}^2$.

Next, we compare 3 cases with different white-noise levels (1 ns, 10 ns and 100 ns), while fixing the distance precision to 1 pc.
The right panel of \cref{fig:deg2} shows the result.
Without pulsars with a precise distance (the leftmost value), the white-noise level largely affects the angular resolution.
However, the dependence on the white-noise level becomes weak as the number of pulsars with a precise distance increases.
In fact, for the number larger than 6, the difference is only a factor of 3.
In other words, distance information can reduce the effect of the ToA noise on the localization of GW sources.
We see that, even in a case with 100 ns noise and 1 pc distance uncertainty, which will be practical in the near future, the precision improves by 2 orders of magnitude.
Here, the mean and standard deviation of angular resolution for the noise level of 100 ns are also plotted in the right panel of \cref{fig:deg2}.
These are obtained from 10 simulations with different white-noise realizations in addition to the main simulation.
Although relatively large statistical fluctuations can be seen in angular resolution for the number smaller than 7, the mean is monotonically decreasing with the number, as expected.

\par

\section{Conclusions and Discussions}\label{sec: Conclusion}

We have studied the impact of precise pulsar distance measurements on the pulsar timing array.
If the precise distance of pulsars is given by external observations such as VLBI astrometry, it is expected that the phase of the pulsar term can be determined precisely, and the precision of the determination of the model parameters, especially the sky location of the gravitational wave source, will be improved significantly.
To evaluate the impact, we created timing residual data of 12 pulsars for 12.5 years with Gaussian white noise in the presence of the gravitational wave signal generated by the supermassive black hole binary.
We performed Bayesian analysis to estimate the uncertainties of the parameters, incorporating external information from independent observations by considering priors on the pulsar distance.

First, we considered a case with very small white noise of 1 nsec to see the potential usefulness of precise pulsar distance.
While the uncertainty area was $40~{\rm amin}^2$ for a conservative distance precision of 100 pc, it improved to $2~{\rm asec}^2$ for an extreme precision of 0.01 pc.
Other parameters which are directly related to the pulsar-term phase such as $\mathcal{M}$ and $f_{0}$ were also found to improve significantly.
It was noticeable that only 3 pulsars with precise distances can improve the localization of the gravitational wave source.

Next, we varied distance precision and white-noise level systematically.
We found that even in a currently practical case with 100 ns white noise, if half of the pulsars have a distance precision of 1 pc, the localization can improve by more than 1 order of magnitude compared with a case without external distance information.
Furthermore, in a case with 10 ns white noise and 1 pc distance precision for all pulsars, which will be practical in the SKA era, the localization can improve by 2 orders of magnitude and the uncertainty area reaches 10 $\rm{amin}^2$.
If even only a few pulsars have a distance precision of 0.1 pc, the improvement is much more drastic.

In this paper, only white noise was considered, and red noise, frequency-dependent noise and other systematic errors were not taken into account.
In practice, various systematic errors can deteriorate or bias the localization of GW sources, which will potentially degrade the results presented here.
We also did not consider the presence of the stochastic GW background.
It would be important to consider how the presence of the stochastic GW background affects the continuous GW search.

Furthermore, we assumed, for simplicity, that pulsars are isotropically distributed on the sky and have a distance of 1 kpc.
The covariance of $\theta$ and $\phi$ and the precision of the localization are considered to depend on the sky distribution of the pulsars and the relative position of the pulsars with precise distance and the GW source.
Also, as was shown in \cite{Goldstein_2018}, the precision of GW source localization depends on whether the GW source is located in a sky region where pulsars are densely distributed or not.
Although considering the realistic pulsar distribution is beyond the scope of the current paper, we present some analyses with two simple cases in \cref{Appendix non-uniform}.
More realistic analyses taking into account systematic errors and 3-dimensional pulsar distributions will be our future work.

\acknowledgments
We thank the referee for helpful comments and suggestions.
KT is partially supported by JSPS KAKENHI Grant Numbers 20H00180, 21H01130, and 21H04467, Bilateral Joint Research Projects of JSPS, and the ISM Cooperative Research Program (2023-ISMCRP-2046).

\appendix
\section{Direction of correlation}\label{Appendix direction}

\begin{figure}[tb]
    \centering
    \includegraphics[width=18cm]{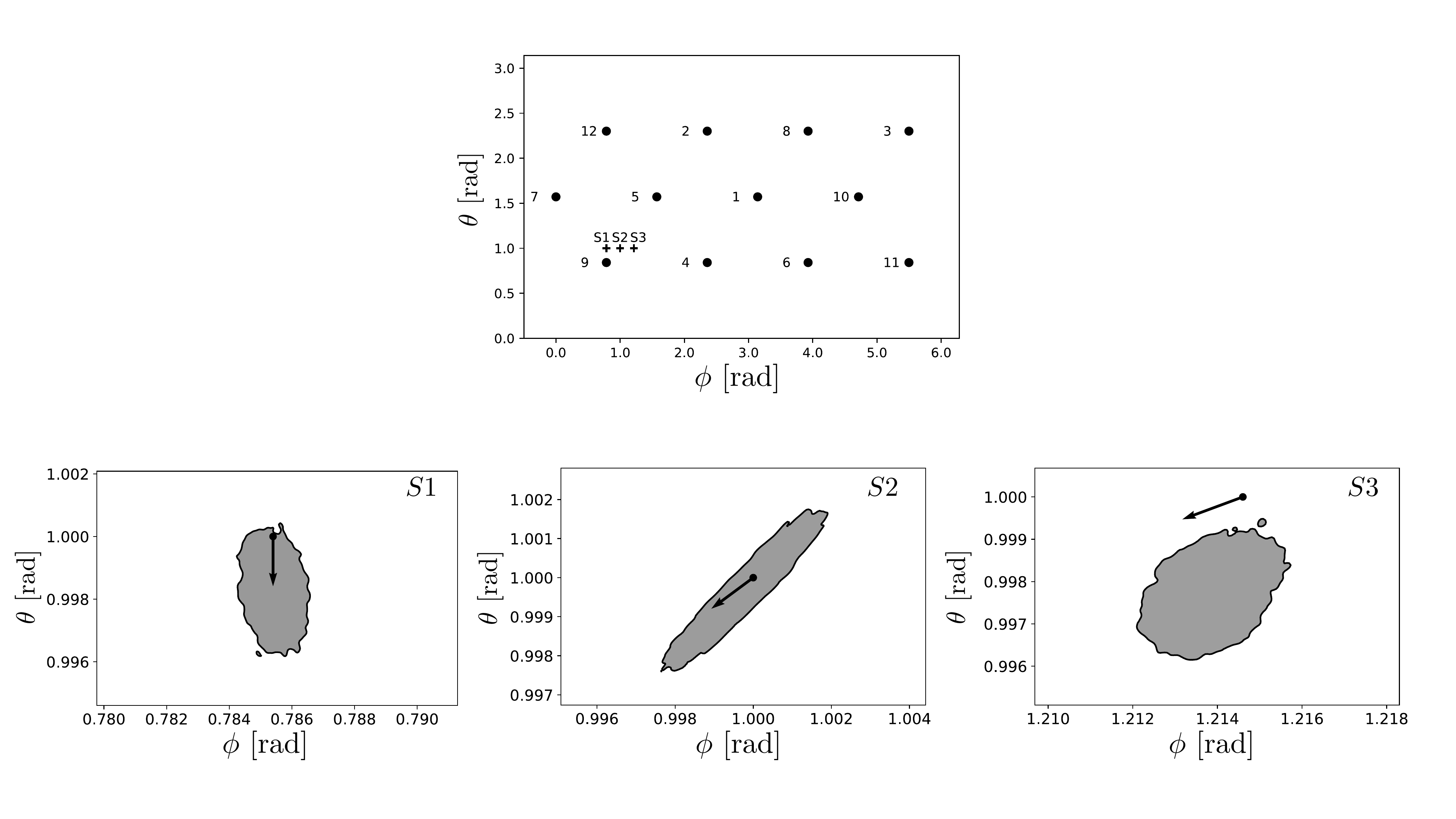}
    \caption{{\it Upper}: Same as the left panel of \cref{fig:posdist} but the plus markers with S1, S2, and S3 denote the near, middle, and far GW sources, respectively.
            {\it Lower}: Posterior of the sky location of the GW sources.
        The left, middle and right panels for the GW sources of S1, S2 and S3, respectively.
        The initial point of the vector denote the true values and the vector directed to the pulsar of the number 9}
    \label{fig:direction}
\end{figure}

Here, to investigate the direction of the correlation of the GW-source sky location, we perform additional analyses with different source locations.
We fix the white-noise level and pulsar distance uncertainty to 1 ns and 100 pc, respectively.
Upper panel of \cref{fig:direction} shows the sky locations of the pulsars and the GW sources.
We consider two additional cases of the GW sky location, which is closer (S1) and farther (S3) to the 9th pulsar compared to the fiducial location (S2) in the main text.

Lower panel of \cref{fig:direction} shows the posterior of the sky location of the GW source.
The direction to the nearest pulsar (No.9) is also shown.
It is seen that the direction of credible region is toward the nearest pulsar.
On the other hand, there is no improvement with respect to the uncertainty area.

\par

\section{Non-uniform distribution of pulsars}\label{Appendix non-uniform}

\begin{figure}[tb]
    \centering
    \includegraphics[width=18cm]{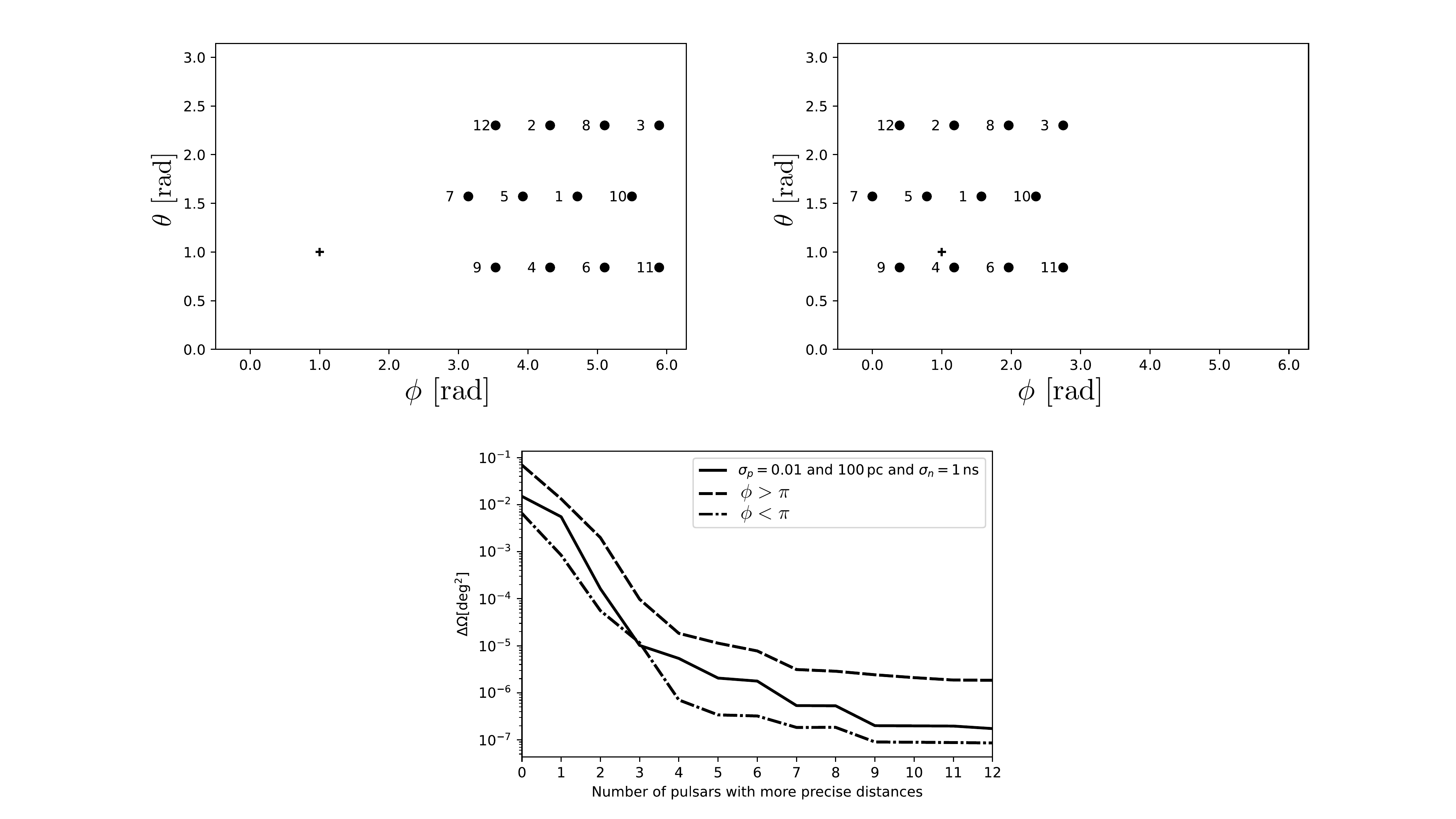}
    \caption{{\it Upper}: Same as the left panel of \cref{fig:posdist} but the pulsars are distributed only in $\phi>\pi$ (left panel) and $\phi<\pi$ (right panel).
            {\it Lower}: Same as \cref{fig:tp} but the dashed line for the case of $\phi>\pi$ and the dashed-dotted line for the case of $\phi<\pi$.
        The solid line is the same line as in \cref{fig:tp}.}
    \label{fig:non-uniform}
\end{figure}

Here, we investigate the effect of non-uniform distribution of pulsars with two simple cases.
The pulsars are distributed only in a region of either $\phi>\pi$ or $\phi<\pi$, fixing the position of the GW source at $\phi = \theta = 1$, as shown in the upper panels of \cref{fig:non-uniform}.
The former (latter) case simulates a situation where a GW source is located where pulsars are sparse and dense, respectively.
Using these distributions and 1 ns white noise, we consider a case in which the distance of only some of the 12 pulsars is precisely measured with a precision of 0.01 pc, while others have a precision of 100 pc.

The lower panel of \cref{fig:non-uniform} shows the angular resolution of the GW source as a function of the number of pulsars with precisely determined distances.
The case of $\phi>\pi$ (dashed) was found to be one or two orders of magnitude larger than the case of $\phi<\pi$ (dashed-dotted).
Then, the fiducial case of uniform distribution (solid) is between these two curves.
This suggests that the pulsars with a precise distance are very effective to improve the uncertainty area, even if they are located far from the GW source.

\par

\bibliography{CWbib_referee}

\end{document}